\documentclass[a4paper,
               ]{jacow}
\usepackage{pdfpages,multirow,ragged2e,subcaption}
\usepackage{hyperref} 
\makeatletter%
	\ifboolexpr{bool{xetex}}
	 {\renewcommand{\Gin@extensions}{.pdf,%
	                    .png,.jpg,.bmp,.pict,.tif,.psd,.mac,.sga,.tga,.gif,%
	                    .eps,.ps,%
	                    }}{}
\makeatother

\ifboolexpr{bool{xetex} or bool{luatex}} 
 {}                                      
 {\usepackage[utf8]{inputenc}}           

\usepackage[USenglish]{babel}

\ifboolexpr{bool{jacowbiblatex}}%
 {%
  \addbibresource{jacow-test.bib}
  \addbibresource{biblatex-examples.bib}
 }{}
\begin{document}

\title{Preliminary results from the CLEAR \\ nonlinear plasma lens experiment}


\author{P. Drobniak\thanks{pierre.drobniak@fys.uio.no}, E. Adli, H. B. Anderson, K. N. Sjobak, C. A. Lindstr{\o}m, \\ Department of Physics, University of Oslo, Oslo, Norway \\
A. Dyson, University of Oxford, Clarendon Laboratory, Oxford, United Kingdom \\
S. M. Mewes, M. Th{\'e}venet, Deutsches Elektronen-Synchrotron DESY, Hamburg, Germany}
    
\maketitle

\begin{abstract}
   Plasma lensing provides compact focusing of electron beams, since they offer strong focusing fields (kT/m) in both planes simultaneously. This becomes particularly important for highly diverging beams with a large energy spread such as those typically originating from plasma accelerators. The lens presented here is a nonlinear active plasma lens, with a controlled focusing-strength variation purposely introduced in one transverse direction. This lens is a key element of a larger transport lattice, core of the ERC project SPARTA, which aims to provide a solution for achromatic transport between plasma-accelerator stages. We report on preliminary experimental results from the CLEAR facility at CERN, which aims to probe the magnetic field structure of the lens using an electron beam, in search of the desired nonlinearity, together with 2D plasma simulation results.
\end{abstract}

\section{INTRODUCTION}

Plasma accelerators \cite{tajima1979} offer accelerating gradients that are several orders of magnitude greater than with RF technology, typically >~GV/m compared to tens of MV/m \cite{HoganPRL2005,LeemansNatPhys2006}. They may be the solution for next hundreds-of-GeV machines such as the HALHF project~\cite{Foster2023,Foster2025}, potentially even opening the possibility to reach multi-TeV ranges.


Whether for laser- or beam-driven plasma acceleration, the limit in energy gain \cite{picksley2024,blumenfeld2007} comes from the driver, which either loses energy or undergoes propagation issues. One solution for reaching higher energies is to couple several plasma accelerators, in order to renew the driver from stage to stage. 
For laser-driven acceleration, this was realised by Steinke et al. \cite{steinke2016}, where two independent plasma stages were coupled by means of an active plasma lens. 
The latter offers azimuthally symmetric and compact focusing for highly diverging beams, such as those from plasma sources and accelerators. In this experiment, the charge-coupling efficiency was, however, limited to a few percent, largely due to the chromaticity of the setup. 

The ERC-funded project SPARTA \cite{spartaProject2023,spartaIPAC2025} proposes to solve this chromaticity issue with a novel achromatic transport lattice \cite{drobniak2025}, composed of dipoles, nonlinear active plasma lenses, and a sextupole. Inspired by nonlinear optics used in collider final-focus systems \cite{raimondi2001}, the nonlinear plasma lenses provide stronger focusing on, say, the right side and weaker focusing on the left side, which when combined with a dispersive dipole can be used to provide stronger focusing to higher-energy particles and weaker focusing to lower-energy particles. This enables the achromatic beam transport between stages and therefore beam-quality preservation. The present article focuses on the theoretical and experimental development of these nonlinear plasma lenses.



We first introduce the concept of active nonlinear plasma lens. Secondly, we describe our first experimental results aimed at characterising the lens. Finally, we present 2D hydrodynamic simulations that motivate the changes planned for the next experimental campaign.

\section{NONLINEAR PLASMA LENS CONCEPT}
As opposed to quadrupoles, plasma lenses offer a compact solution with radial focusing gradient in the kT/m range, orders of magnitude above conventional techniques. The word ``active'' refers to the technique used, whereby a gas is ionised by a high-voltage source, creating a kA current in a discharge capillary, in our case 1~mm wide and 20~mm long. 
Through $\nabla \times \vec{B} = \mu_0 \vec{J}$, the current generates an azimuthal $B$-field that is focusing for an incoming electron beam.



For a negatively charged beam propagating towards negative $z$, the theoretical nonlinear $B$-field distribution we are aiming for in the $xy$-plane of the lens is described by:
\begin{align}
B_x^{\mathrm{nonlin}} &= -g_0 \left( y + \frac{1}{D_x} xy \right) \label{eq:B_nonlin_x} \\
B_y^{\mathrm{nonlin}} &= +g_0 \left( x + \frac{1}{D_x}\frac{x^2 + y^2}{2} \right) \label{eq:B_nonlin_y},
\end{align}
where $g_0$ is the longitudinally averaged focusing gradient, and the (inverse) nonlinearity coefficient is matched to the dipole-induced dispersion $D_x$ from upstream of the lens.


Our solution to create such an asymmetric field is to use a vertical external $B$-field (along the $y$-axis) which acts on the plasma so that the current density rearranges asymmetrically in the $xy$-plane. The main effects we intend to trigger are Hall effect and plasma magnetisation. The first one is motivated by a 1981 article by Kunkel \cite{kunkel1981} where the Lorentz force from an external magnet on a glow discharge creates an asymmetry in the plasma density, while the second refers to the effect of value and orientation of the total $B$-field on plasma properties, such as electrical conductivity \cite{davies2021}. 

\section{PRELIMINARY EXPERIMENTAL RESULTS}
This section presents preliminary results from an experimental campaign at CLEAR \cite{Gamba2018} in 2024.
The operation, as described by Sjobak \textit{et al.} \cite{sjobak2021} consists of measuring the deflection of on incoming electron beam onto a screen downstream of the lens, and using this to reconstruct the $B$-field it traveled through. A full mapping of the $B$-field distribution has not been performed yet, but we present here the search for an obvious nonlinearity, as visible in eq.~\ref{eq:B_nonlin_y}.

Our experiment was performed with argon at a few mbar, injected by pulse at $0.83\,$Hz through a pipe into the lens capillary. Two hollow electrodes ionise the gas at the same frequency, with a $150\,$ns-rise-time $1\,$kA-peak discharge. An external electromagnet creates the vertical external $B$-field. 


The electron beam had the following characteristics: $185 \pm 1\,$MeV energy, $<10\,$pC charge per bunch, 4 bunches per train ($\approx 3\,$ps each, separated by $667\,$ps), $1\,$train every $1.2\,$s ($0.83\,$Hz). It was focused at the lens entrance to $50\times50\,$\textmu m rms. Synchronisation between beam and discharge was ensured, to a precision below $5\,$ns rms.

Alignment and relative-offset scans were performed by moving the lens relative to the beam. Magnets upstream of the lens offered additional tuning of the incoming electron beam. The final position of the outgoing beam was observed on a fixed YAG screen (not moving with the lens), positioned at a distance $\Delta{}z = 30\,$cm downstream of the lens exit. 

For an electron beam traveling towards negative $z$, the desired beam displacements on the screen are of the form (when assuming a thin lens and ultra-relativistic beam):
\begin{align}
\Delta{}x(x_l,y_l) &= -\frac{e c L \Delta{}z}{E} g_0 \left(-x_l+\frac{1}{D_x}\frac{x_l^2+y_l^2}{2}\right)
\label{eq:kick_x} \\
\Delta{}y(x_l,y_l) &= +\frac{e c L \Delta{}z}{E} g_0 \left(y_l+\frac{1}{D_x}x_l y_l\right)
\label{eq:kick_y},
\end{align}
with $(x_l,y_l)$ the lens position with respect to the beam axis, $e$ the electron charge, $c$ the speed of light, $L$ the length of the lens and $E$ the beam energy. The direct kick from the dipole external $B$-field is ignored, since it generates a constant offset. 

In order to observe clear evidence of a nonlinear lensing effect on the screen, we decided to perform a vertical offset scan across the lens through its magnetic center, with $x_l = 0$.
The vertical displacements $\Delta{}y$ are linearly growing with $y_l$ while the horizontal displacements $\Delta{}x$ are purely quadratic. This results in a ``banana''-like shape on the screen, flipping direction when reversing the nonlinearity, which is expected to depend on the polarity of the external magnetic field.

\begin{figure}[t]
    \centering
\includegraphics[width=1.0\linewidth]{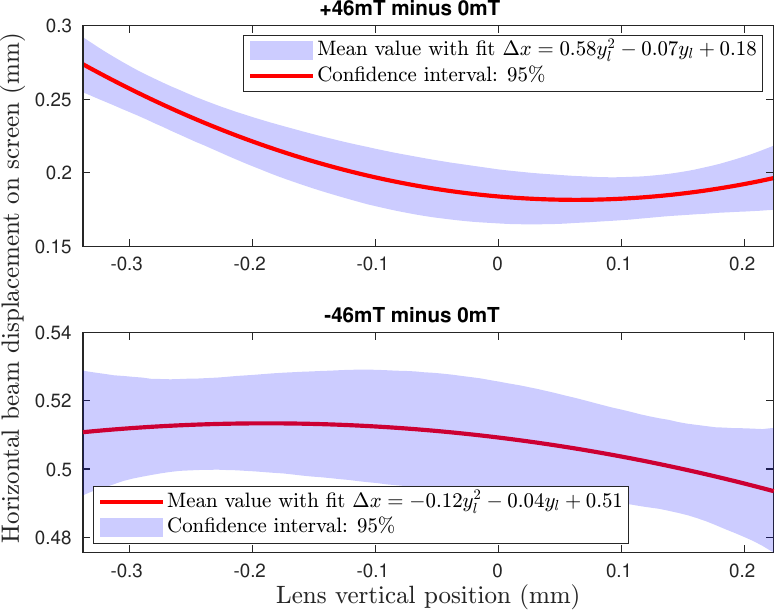}
    \caption{Prelim. exp. results of horizontal beam displacements observed on the screen from vertical scans with an external field of $+46\,$mT (top) and $-46\,$mT (bottom), relative to the displacements from a vertical scan without external field. The dipole contribution from the external field has been removed. The lens was centered horizontally.}
    \label{fig:banana_exp}
\end{figure}

The lens was translated in steps of $50\,$\textmu m. For each $y_l$, the beam centroid positions are recorded by averaging over 10 consecutive shots. Lens positions yielding beams with too much scattering are excluded. The theoretical full $1$~mm aperture was thus reduced to \SI{\sim0.6}{mm} due to angular misalignments. 
We performed three scans with discharge: with external field at $+46\pm1\,$mT ($B$-field pointing up) and at $-46\pm1\,$mT ($B$-field pointing down), and without external field. 
In addition, we record the effect of the same external fields ``only'' by switching off the discharge.
Finally, to isolate the nonlinear effect, we subtract the third scan (i.e., discharge without external field) from the first and second scan.
For each lens position, we retrieve the mean beam displacement, and create the corresponding Gaussian distribution. We then generate $10000$ ``virtual scans'', with beam displacements randomly picked in each position normal distribution. This Monte Carlo approach results in a fit, with corresponding confidence intervals, as presented in Fig.~\ref{fig:banana_exp}. 


We can indeed see signs of a ``banana''-shape, which reverses when we flip the magnet polarity. Values are not centered around $0$, which may originate from a lens misalignment. However, the key measurement of the curvature is, to first order, not affected by this misalignment. The coefficient of curvatures are $m_2 = +0.57$ and \SI{-0.12}{\per\mm} for the +46 and \SI{-46}{mT} cases, respectively. Inverting Eq.~\ref{eq:kick_x} for $x_l=0$, we find that the nonlinearity is given by 
\begin{equation}
    \frac{1}{D_x} = \left(\frac{2 E}{e c L \Delta z g_0}\right) m_2\,,
\end{equation}
where $g_0 = \mu_0 I_0 / (2\pi{}R^2) = 800$~T/m with $I_0 = \SI{1}{kA}$, $L$ is \SI{20}{mm}, $\Delta z$ is \SI{300}{mm}, and $E$ is \SI{185}{MeV}. The best-fit nonlinearities are therefore $\left(1/D_x\right)^{+\SI{46}{mT}} = \SI{+146}{\per\m}$ and $\left(1/D_x\right)^{-\SI{46}{mT}} = \SI{-31}{\per\m}$. Treating this as a difference measurement, which should remove any systematic curvature offset, we can take the average of the above, resulting in $\left|1/D_x\right|^{\mathrm{average}} = \SI{89}{\per\m}$, which is close to the desired range of values around $100$~m$^{-1}$ as defined for the SPARTA project. Note, however, that the error bars are large: we cannot from this data exclude the possibility of there being no effect.


The sign of the curvature coefficient is not yet well understood from theoretical considerations. From a naive treatment of the Lorentz force on the plasma electrons (pushing the discharge to one side), and assuming a constant plasma pressure, we would deduce the opposite sign for the curvature coefficient compared to the observation. However, considering instead the effect of magnetisation on the plasma conductivity, we would indeed expect the observed sign of the coefficient. Hydrodynamic simulations, including magnetisation, are required to disentangle these effects.


It is not clear to us at this point how badly the misalignment influenced the measurement. A more precise and tunable plasma-lens design, already constructed for an upcoming campaign, will hopefully unlock more accurate measurements. Accounting for thick lens effects in the analysis will also help. Further, we will increase the strength of the external magnetic field in order to probe stronger plasma interactions---this should make the effect more visible.


\section{HYDRODYNAMIC SIMULATION}
Our goal in this section is not yet to reproduce the CLEAR experiment, since the model we employ uses H$_2$, which does not behave similarly to a heavier gas like Ar, as experimentally observed by Lindstr{\o}m \textit{et al.}~\cite{Lindstrom2018}. Instead, our goal is to extend our previous 1D study (see Ref.~\cite{drobniak2025}) to 2D-simulations, in order to better understand the origin of the nonlinearity. The present model runs in COMSOL, developed at DESY \cite{mewes2023} and originally intended to study H$_2$ interactions, integrates the effect of the $B$-field (both internal and external field) on the electrons through the Lorentz force and takes plasma magnetisation into account \cite{davies2021}. Work is currently ongoing to extend this model to Ar.

\begin{figure}[htbp]
    \centering
    \includegraphics[width=1.0\linewidth]{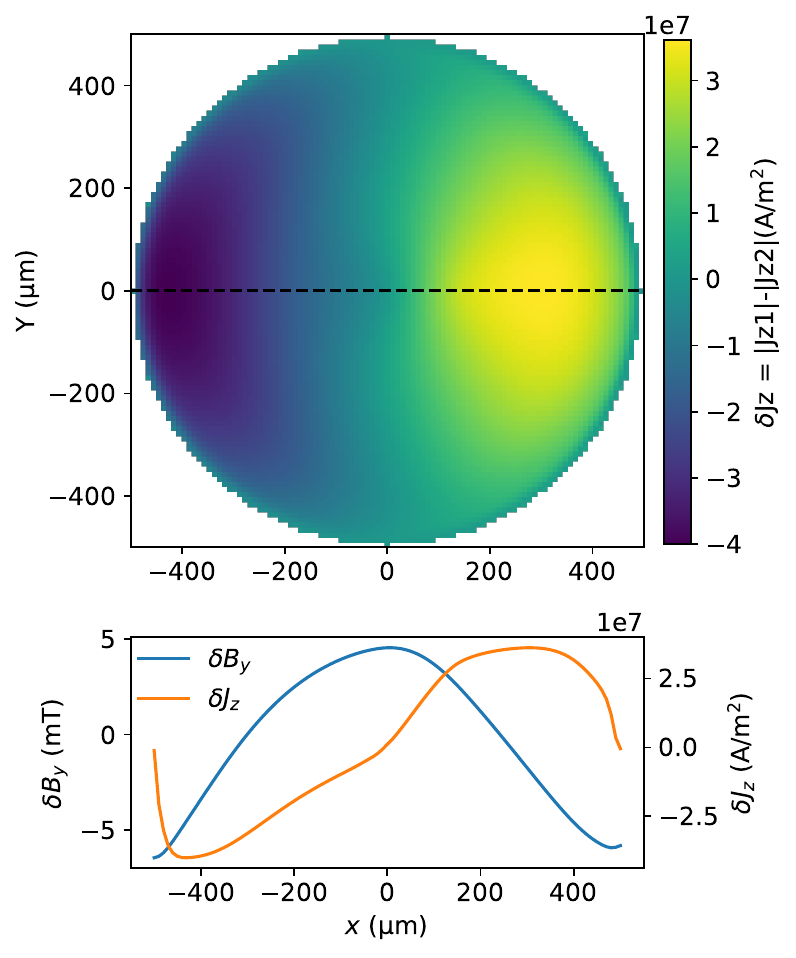}
    
    \caption{Top: difference in 2D absolute current density $\delta{}J_z$ between a simulation with $160\,$mT external field and no external field, both with H$_2$ at $7\,$mbar, extracted at $185\,$ns. A black dashed line indicates where data are extracted for the bottom graph. Bottom: 1D plots extracted at $(x_l,y_l)=(x,0)$ for the difference in vertical $B$-field $\delta{}B_y$ (where the external $B$-field has been subtracted) and in current density.}
    \label{fig:By_current_2D}
\end{figure}

Our simulations are performed with the same geometry as the experiments, as well as the same current profile. The gas is initialised as H$_2$ at $7\,$mbar, slightly ionised. The external field is $160\,$mT (3.5 times higher than in the experiment described above). 
A typical 2D-simulation result, for a fully ionised plasma, is presented in Fig.~\ref{fig:By_current_2D}. In order to virtually suppress the symmetrical nonlinearity caused by wall-cooling effects (present in H$_2$), making it more indicative of what is achievable in a gas without this effect (such as Ar \cite{Lindstrom2018}), we plot the difference in absolute current density between a simulation with \SI{160}{mT} external field and a simulation without external field: $\delta J_z = |(J_z)^{\SI{160}{mT}}|-|(J_z)^{\SI{0}{mT}}|$.


The difference in current density distribution is clearly asymmetric, implying that the external field indeed has an effect. This is also visible in the internal vertical $B$-field difference along the horizontal axis ($\delta{}B_y (x,0)$), which largely has a quadratic shape (but also an offset, which indicates a small shift of the magnetic center). In order to quantify the nonlinearity, we performed a third-order polynomial fit on $\delta{}B_y$ in the range $[-390,390]\,$\textmu m. 
The present simulation has $(g_0)_{\mathrm{sim.}} = 1130\,$T/m and $\left(1/D_x\right)_{\mathrm{sim.}} = \SI{100}{\per\m}$, which again is very encouraging as it overlaps with the SPARTA requirements. Some higher order terms are also present: we are aiming to remove these by simulating a heavier gas (Ar) in the near future. 

Plasma magnetisation only plays a marginal role here and we observe a strong contribution from the Lorentz force on the electron density. This motivates the continuation of our CLEAR experimental campaign coupled with the development of a heavy-gas model for a better understanding of our nonlinear lens concept.

\section{CONCLUSION}
The concept of a nonlinear plasma lens for achromatic staging has been introduced, with its theoretical required $B$-field distribution. Our first experimental results (with Ar) from an experimental campaign at CLEAR in 2024 show a potential nonlinearity. Simulations in 2D, but with a lighter gas (H$_2$), appear to confirm that our concept could work, motivating a new experimental campaign with better alignment and stronger external magnetic fields. 

\section{ACKNOWLEDGEMENTS}

The authors wish to thank J.~S. Ringnes, S.~R.~Solbak, H.~Borg at the I-Lab, Uni.~Oslo and R.~Corsini, W.~Farabolini, A.~Gilardi, P.~Korysko, V.~Rieker at CLEAR, CERN. This work is funded by the European Research Council (ERC Grant Agreement No. 101116161), the Research Council of Norway (NFR Grants No. 313770, 310713 and 353317) and supported in part through the Maxwell computational resources operated at Deutsches Elektronen-Synchrotron DESY, Hamburg, Germany.

%
%
\ifboolexpr{bool{jacowbiblatex}}%
	{\printbibliography}%
	{%
	
	
    } 
%
%


\end{document}